\documentclass{iau}
\usepackage{graphicx}

\title[Pulsars as probes for the Galactic magnetic fields] %% short title %%
{Pulsars as excellent probes\\ for the magnetic structure in our Milky Way} %% full title %%

\author[J.L. Han]  %% short author list %%
{JinLin Han %$^1$
}

\affiliation{
National Astronomical Observatories, Chinese Academy of Sciences, 
\\ DaTun Road 20A, ChaoYang District, Beijing 100012, China 
\\ email: {\tt hjl@nao.cas.cn} 
}

\pubyear{2012}
\volume{291}  %% insert here IAU Symposium No.
\jname{\mbox{Neutron Stars and Pulsars: Challenges and Opportunities after 80 years}}
\editors{J. van Leeuwen, ed.} 
\begin{document}

\maketitle

%\baselineskip=17pt

%% -- Abstract ----------------------------------
\begin{abstract}
In this invited talk, I first discuss the advantages and disadvantages
of many probes for the magnetic fields of the Milky Way. I conclude
that pulsars are the best probes for the magnetic structure in our
Galaxy, because magnetic field strength and directions can be derived
from their dispersion measures (DMs) and rotation measures
(RMs). Using the pulsars as probes, magnetic field structures in the
Galactic disk, especially the field reversals between the arms and
interarm regions, can be well revealed from the distribution of 
RM data. The field strengths on large scales and small scales can be
derived from RM and DM data. RMs of extragalactic radio sources can be
used as the indication of magnetic field directions in the spiral
tangential regions, and can be used as probes for the magnetic fields
in the regions farther away than pulsars when their median RMs are
compared with pulsar RMs.
\end{abstract}

\firstsection % if your document starts with a section,
\section{Introduction}

Magnetic fields permeate the interstellar medium on all scales, from
the stellar scale of AU size to the galactic scale of tens of kpc. In
general, magnetic fields are tangled by many physical process in the
interstellar medium, for example, the localized processes of star
formation and supernova explosion, the galactic processes of
differential rotation, density wave and stream motion, so that the
ever-existed large-scale magnetic fields are very distorted. When we
try to understand the properties of magnetic fields in our Milky Way,
we have to know the magnetic fields of different scales in different
regions.

There are many probes for the magnetic fields in our Milky Way. The
most widely used are the starlight polarization, the polarized
emission of dust and clouds at millimeter and submillimeter
wavelength, the Zeeman effect of spectral lines or maser line from
clouds or clumps, the diffuse radio synchrotron emission from
relativistic electrons in the interstellar magnetic fields, the
Faraday rotation of extragalactic radio sources and pulsars. The first
three are related to magnetic field in clouds, and the later two are
related to the fields in the diffuse medium. Each probe measures only
one of the three dimensional field components or their average or
integration. We have to understand the advantages and disadvantages of
these probes, and then ``connect'' all measurements to get the real
structure of the Galactic magnetic fields.

\section{Advantage and disadvantage of different probes for magnetic fields}

Some of of the above mentioned probes are more suitable to reveal the
localized magnetic fields, and some are more sensitive to the
large-scale fields. Because of our location at the disk edge in the
Milky Way, the best probes for the Galactic magnetic structure have to
be able to detect the magnetic component parallel to the line of
sight.

\subsection{Starlight polarization}

Starlight is not polarized when it is radiated. It becomes polarized
when it passes through the interstellar medium. It is slightly
absorbed or scattered by interstellar dust grains which are
preferentially aligned by interstellar magnetic fields. The farther
the star is, the more extinction its light suffers on the path to us,
and the higher the polarization of starlight. Starlight polarization
is the measurement of the summed extinction due to dusts in the path,
therefore it indicates the averaged magnetic field orientations in the
sky plane, not directions.

The polarization of stars around an individual cloud can be used to
trace the magnetic field orientation inside the cloud.  When more and
more stars are observed in the very deep Galactic disk, the data can
not separate magnetic field orientations in every clouds at different
distances which on average are generally aligned with the Galactic
plane. Therefore, the starlight probes can not be used for the
large-scale magnetic structure in the Galactic disk. They are good
probes for the local halo fields if a large number of high-latitude
distant stars are observed in wide sky area.

%\vspace{-1mm}
\subsection{Polarized emission of dust and clouds}

The polarized emission from the clouds or dust is quite localized
feature, and can show the magnetic field orientations in the clouds
perpendicular to the line of sight. There is evidence that the
magnetic fields in clouds are related to the fields of larger scales
(\cite{lh11}).  Similar to the starlight probes, even if the magnetic
orientations of many clouds in the disk of the Milky way are observed,
we cannot get the large-scale magnetic structure, because the field
orientation of all clouds are aligned with the Galactic plane in
general. In fact, the polarized emission of dust and clouds has
already been seen clearly from the polarization maps of WMAP or
Planck, which indicate such parallel fields to the Galactic plane.

%\vspace{-1mm}
\subsection{The Zeeman effect of spectral lines from clouds or clumps}

The line emission or absorption from the clouds or clumps with
internal magnetic fields show Zeeman splitting of the line. The
separation of the split lines measures the field strength of the
magnetic component parallel to the line of sight, and the sense change
of circular polarization of the split lines indicates the direction of
the field component on the line of sight. Physically the measurements
are sensitive only to the fields inside the clouds or clumps, which
are of a scale of pc or even AU. Surprisingly, the distribution of
available magnetic field measurements from maser lines around HII
regions are very coherent with the large-scale spiral structure and
large-scale magnetic fields (\cite{hz07}). A big project using the
ATCA (\cite{gmc+12}) is in progress for extensive observations of the
Zeeman splitting of lines from HII regions for large-scale magnetic
structure.

%\vspace{-1mm}
\subsection{Diffuse radio synchrotron emission and polarization}

The polarization of synchrotron radiation from relativistic electrons
shows the average orientation of magnetic fields in the sky plane in
the emission region, perpendicular to the line of sight. If random
magnetic fields dominate in the emission region, as is often the case, then
polarization ``vectors'' (orientation, not direction) could be very
random, and the observed emission is depolarized. The polarized
emission from different regions inside the Milky Way at different
distances also suffers different Faraday rotations when it propagates
in the interstellar medium. As we are located on the edge of the disk
of the Milky Way, the observed synchrotron radiation is the
superposition of such variously Faraday-rotated polarized radiation
from everywhere at all distances until the Milky way ``boundary'',
which could be therefore very depolarized. The lower the observational
frequency, the stronger the depolarization. The closer towards the
Galactic Center along the Galactic plane, the less ordered the
polarization. In the outer region near the anticenter of the Milky
way, the radio emission comes from only the Perseus arm, and we see
more polarized emission than in the inner Galaxy (\cite{xhr+11}).

Using the polarization survey of radio synchrotron of the whole sky,
one can get some constraints on the magnetic fields in the halo of Milky
Way, but it is hard to constrain the magnetic fields in the Galactic
disk. Because there is no linearity for Faraday rotation against
distance, and because the similarly polarized emission can come from
regions of different distances, the rotation measure synthesis cannot 
separate the polarized emission from different regions.

%\vspace{-1mm}
\subsection{Rotation measures of extragalactic radio sources}

Faraday rotation is the summed rotation of the polarization angle
$\phi$ of a linearly polarized wave on the way from a source
($\phi_0$) to us, $\phi=RM*\lambda^2+\phi_0$. The rotation measure
(RM) is related to electron density and magnetic fields by ${\rm RM} =
0.810 \int_{\rm source}^{\rm observer} n_e {\bf B} \cdot d{\bf l}$ (in
unit of rad~m$^{-2}$). Here $n_e$ is the electron density in
cm$^{-3}$, ${\bf B}$ is the vector magnetic field in $\mu$G and $d
{\bf l}$ is an elemental vector along the line of sight toward us in
pc. Obviously, the RM is sensitive to the magnetic field component on
the line of sight, weighted by the electron density. Notice that it is
an integrated value. Positive RMs correspond to the average fields on
the path directed toward us. Random or smaller-scale fields on the
path cannot be recognized from the final observed wavelength dependence 
of $\phi$.

There are many background extragalactic radio sources (EGRs)
distributed in the sky, which can be the most powerful probe of the
magnetic field in the halo of our Milky Way. The observed RMs contain
three contributions: 1) the intrinsic RM from the source, which
depends on the observational wavelength and how deep in the source
the radiation comes from, 2) the RM of intergalactic medium
which is probably very small and not measured yet, and 3) the RM
foreground from our Milky Way. The RMs of a number of EGRs in a given
sky region should have more or less the common foreground Galactic RM
contribution. The large-scale RM sky distribution therefore shows the
magnetic fields in the halo (\cite[Han et al. 1997, 1999]{hmbb97,hmq99}). The RM sky of more
dense data set (\cite{tss09, ojr+12}) can show the details of magnetic
fields in visually large objects (\cite{hmg11}), in addition to the
general RM sky of large angular scales.

At lower Galactic latitudes, extensive efforts have recently been
made to enlarge the RM samples (e.g. \cite{bhg+07,vbs+11}). Towards
the central region the data become more scarce, because the
diffuse emission is stronger and because the polarization
observations are more difficult to carry out. The median RMs of
background EGRs behind the disk are the integrated measurement of
polarization angle rotations over the whole path in the disk and
therefore not sensitive to the possible magnetic field reversals
between arms and interarm regions on the path. The dominant
contribution to RMs of EGRs comes from tangential regions, where
the magnetic fields have the smallest angle with the line of sight if
the fields follow spiral arms. When a set of RMs of EGRs are fitted
with a magnetic structure model, the electron density model is a
necessary independent input.

\begin{figure}
\centering\includegraphics[angle=270,width=0.98\textwidth]{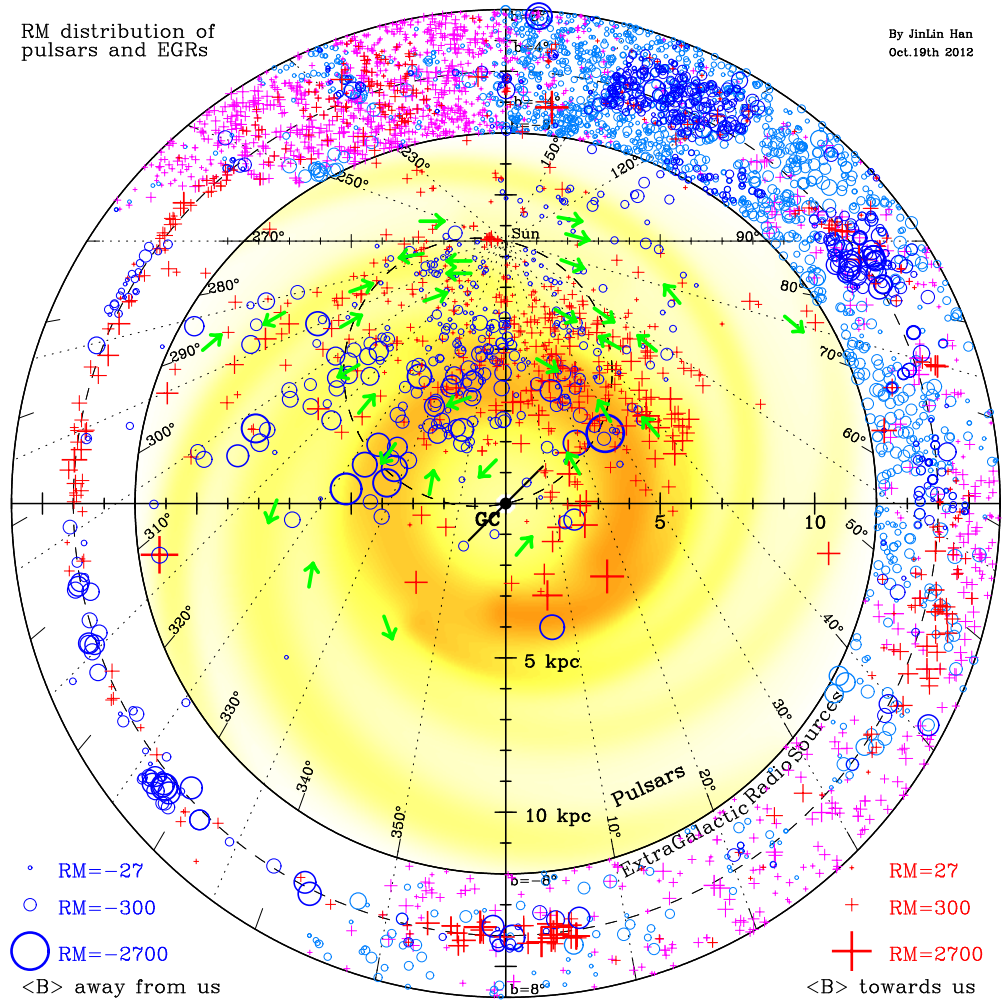}
\vspace{3mm}
\caption{The RM distribution of 736 pulsars of $|b|<8^{\circ}$
  projected onto the Galactic plane, including new data of Han et al
  (2012, in preparation). The linear sizes of the symbols are
  proportional to the square root of the RM values with limits of
  $\pm$27 and $\pm$2700 rad~m$^{-2}$. Positive RMs are shown by plus
  signs and negative RMs by open circles. The background shows the
  approximate locations of spiral arms used in the NE2001 electron
  density model (\cite{cl02}). Published RMs of EGRs of
  $|b|<8^{\circ}$ are displayed in the outer ring according to their
  $l$ and $b$, with the same convention of RM symbols and limits. The
  data from the NVSS RM catalog (\cite{tss09}) are plotted in
  light-blue and pink symbols. The large-scale structure of magnetic
  fields in the Galactic disk, as indicated by arrows, are derived
  from the distribution of pulsar RMs and the comparison of pulsar RMs
  with the RMs of background EGRs.
\label{rm_psr_egr}
}
\end{figure}

\vspace{-2mm}
\section{Pulsars as probes for the Galactic magnetic fields}

Pulsars are polarized radio sources inside our Milky Way. The observed
RMs of pulsars come only from the interstellar medium between pulsars
and us, because there is no intrinsic Faraday rotation from the
emission region and pulsar magnetosphere (\cite{whl11}). For a pulsar
at distance $D$ (in pc), the RM is given by
$ %\begin{equation}
{\rm RM} = 0.810 \int_{0}^{D} n_e {\bf B} \cdot d{\bf l}.
$ %\end{equation}
With the pulsar dispersion measure, 
$ %\begin{equation}
{\rm DM}=\int_{0}^{D} n_e d l,
$ %\end{equation}
we obtain a direct estimate of the field strength weighted by the local free
electron density
\begin{equation}\label{eq_B}
\langle B_{||} \rangle  = \frac{\int_{0}^{D} n_e {\bf B} \cdot d{\bf
l} }{\int_{0}^{D} n_e d l } = 1.232 \;  \frac{\rm RM}{\rm DM}.
\label{eq-B}
\end{equation}
Pulsars are spread through the Galaxy at approximately known
distances, allowing three-dimensional mapping of the magnetic
fields. If pulsar RM data are model-fitted with the magnetic field
structures with the electron density model (\cite{hq94}), then the
pulsars and EGRs are more or less equivalently good as probes for the
magnetic structure. But when RM and DM data are available for many
pulsars in a given region with similar lines of sight, e.g. one pulsar
at $d0$ and one at $d1$, the RM change against distance or DM can
indicate the direction and magnitude of the large-scale field in
particular regions of the Galaxy (\cite[Han et al. 1999, 2002, 2006]{hmq99,hmlq02,hml+06}). Field
strengths in the region can be directly derived by using
\begin{equation}
\langle B_{||}\rangle_{d1-d0} = 1.232 \frac{\Delta{\rm RM}}{\Delta{\rm DM}}, 
\label{delta_rm_dm}
\end{equation}
where $\langle B_{||}\rangle_{d1-d0}$ is the mean line-of-sight field
component in $\mu$G for the region between distances $d0$ and $d1$,
$\Delta{\rm RM} = {\rm RM}_{d1} - {\rm RM}_{d0}$ and $\Delta{\rm DM}
={\rm DM}_{d1} - {\rm DM}_{d0}$. Notice that this derived field is
not dependent on the electron density model. 

As shown in Han et al. (2006)\nocite{hml+06}, the available pulsar RM
data show that magnetic fields in the spiral arms (i.e. the Norma arm,
the Scutum and Crux arm, and the Sagittarius and Carina arm) are
always counterclockwise in both the first and fourth quadrants, though
some disordered fields appear in some segments of some arms. At least in
the local region and in the fourth quadrant, there is good evidence
that the fields in interarm regions are similarly coherent, but
reversed to be clockwise. Therefore at least four or five reversals in
the fourth quadrant occur from the centre to the outskirts of our
Milky Way. In the central Galactic region interior to the Norma arm,
new RM data of pulsars indicate that the fields are clockwise,
reversed again from the counterclockwise field in the Norma arm. In
the first Galactic quadrant, because the separations between spiral
arms are so small, the RM data are dominated by counterclockwise
fields in the arm regions though a few negative pulsar RMs
indicate clockwise fields in the interarm regions.

We notice that the averaged variation of RMs of extragalactic radio
sources along the Galactic longitudes (\cite{bhg+07}) are consistent
with the field reversal pattern obtained from pulsar RMs.

Using pulsar RM and DM data, Han et al. (2006)\nocite{hml+06} were able
to measure the strength of regular azimuthal fields near the
tangential regions in the 1st and 4th Galactic quadrants. Although the
``uncertainties'', which in fact reflect the random fields, are large,
the tendency is clear that fields get stronger at smaller
Galactocentric radius and weaker in interarm regions. The radial 
variation is, 
\begin{equation}
 B_{\rm reg}(R) = B_0 \; \exp \left[
  \frac{-(R-R_{\odot})} {R_{\rm B}} \right] , 
\end{equation}
with the strength of
the large-scale field at the Sun, $B_0=2.1\pm0.3$ $\mu$G, and the
scale radius $R_{\rm B}=8.5\pm4.7$ kpc, $R$ is the distance from 
the Galactic center, $R_{\odot}=8.5$~kpc is the galactocentric 
distance of the Sun.

Pulsar RMs have also been used to study the small-scale random
magnetic fields in the Galaxy. Some pairs of pulsars close in sky
position have similar DMs but very different RMs, indicating an
irregular field structure on scales of about 100~pc. Some of these
irregularities may result from HII regions in the line of sight to a
pulsar (\cite{mwkj03}).  It has been found from pulsar RMs that the
random field has a strength of $B_r\sim 4-6\mu$G independent of
cell-size in the scale range of 10 -- 100~pc. From pulsar RMs in a
very large region of the Galactic disk, Han et
al. (2004)\nocite{hfm04} obtained a power law distribution for
magnetic field fluctuations of $E_B(k)= C \ (k / {\rm
  kpc^{-1}})^{-0.37\pm0.10}$ at scales from $1/k=$ 0.5~kpc to 15~kpc,
with $C= (6.8\pm0.8)\ 10^{-13} {\rm erg \ cm^{-3} \ kpc}$,
corresponding to an rms field of $\sim6\mu$G in the scale range.

\section{Conclusions and discussions}

Pulsars are excellent probes of the magnetic fields in our Milky Way.
They are the best to reveal the magnetic field structure in the
Galactic disk, especially the field reversals; they are the best to
derive the magnetic field strength -- much less model-dependent than
other probes; and they are the best to get the observational spatial
energy spectrum of the magnetic fields. Pulsars can also be used to
probe the magnetic fields in the Galactic halo.  We already have RMs
for about half of the known pulsars and these have been used for
studies of the Galactic magnetic fields. In the future, when more and more
known pulsars are observed for their rotation measures, we can get
more details of magnetic structure in the nearby half of the Galactic
disk.

Note that magnetic fields in the Galactic disk of the far side of the
Galactic center are not yet explored; very few distant pulsars have
been found there. When more and more pulsars are discovered in the far
half of the disk, using FAST and SKA, for example, we can study the
differences in RMs and DMs of pulsars at various distances in
different arms, and measure the magnetic field directions and strength
in the remote arms, so that the global structure of the disk fields
can be well revealed. However, if the large-scale magnetic fields
always go along arms, as present data suggest, the RMs of distant
pulsars will become less sensitive to the magnetic fields in the far
half disk because the lines of sight will be more perpendicular to the
spiral arms than in the nearby half. The distribution of the magnetic
fields of OH masers in HII and star formation regions can always be
used as supplementary tools for large-scale magnetic fields
(\cite{hz07}), if the large-scale fields of the large-scales are
somehow ``remembered'' in clouds or clumps of small scales as
currently available data suggested.

\begin{acknowledgments}
\noindent Thanks for reading the manuscript go to Dr. Dick Manchester, who has
cooperated with me for many years on the topic. The author is
supported by the National Natural Science Foundation of China
(10833003).
\end{acknowledgments}

\end{document}